\documentclass[aps,pra,twocolumn,superscriptaddress]{revtex4}
\usepackage{amsmath}
\usepackage{amsfonts}
\usepackage{epsfig}
\newcommand{\ket}[1]{\left|{#1}\right>}
\newcommand{\bra}[1]{\left<{#1}\right|}
\newcommand{\floor}[1]{\left\lfloor #1 \right\rfloor}
\begin{document}

\title{Encoding many qubits in a rotor}
\author{Amir Kalev}
\affiliation{Centre for Quantum Technologies, National University of Singapore, 3 Science Drive 2, 117543, Singapore}
\author{Philippe Raynal}
\affiliation{Centre for Quantum Technologies, National University of Singapore, 3 Science Drive 2, 117543, Singapore}
\author{Jun Suzuki}
\affiliation{National Institute of Informatics, 2-1-2 Hitotsubashi, Chiyoda-ku, Tokyo, 101-8430, Japan}
\author{Berthold-Georg Englert}
\affiliation{Centre for Quantum Technologies, National University of Singapore, 3 Science Drive 2, 117543, Singapore}
\affiliation{Department of Physics, National University of Singapore, 2 Science Drive 3, 117542, Singapore}

\date{\today}

\begin{abstract}
We propose a scheme for encoding many qubits in a single rotor, that is, a continuous and periodic degree of freedom. A key feature of this scheme is its ability to manipulate and entangle the encoded qubits with a single operation on the system. We also show, using quantum error-correcting codes, how to protect the qubits against small errors in angular position and momentum which may affect the rotor. We then discuss the feasibility of this scheme and suggest several candidates for its implementation. The proposed scheme is immediately generalizable to qudits of any finite dimension.
\end{abstract}

\maketitle
\maketitle 

\section{Introduction}
Quantum information processing, quantum simulation, and quantum communication require the individual manipulation and the coupling of quantum bits (qubits). Early candidates for qubits were two-dimensional physical systems, like the polarization of a photon or the spin of an electron, in which the state of the system directly represents the state of the qubit. Later, higher dimensional systems like atoms have been considered. There, a qubit is encoded in two energy levels while the remaining dimensions are ignored \cite{cirac95}. Infinite dimensional systems have been studied too. For instance, the encoding of a single qubit (actually qudit, a \mbox{$d$-dimensional} version of a qubit) in continuous variables and harmonic oscillators has been described in \cite{preskill01} and \cite{sanders02}. Importantly, a common feature of all these encodings is the use of a single degree of freedom to encode a {\it single} qubit.

An interesting alternative is to use a single degree of freedom to encode {\it many} qubits. Let us consider an infinite dimensional system which can be decomposed into a qubit subsystem and a system isomorphic to the original system itself, a feature only possible for infinite dimensional systems. Then the same decomposition can be repeated over and over again to obtain  as many qubits as desired. This is the core idea of this article. We show how to encode $N$ qubits (qudits) in a rotor, that is, a single continuous and periodic degree of freedom. The $N$ qubits can be individually manipulated and coupled. Remarkably, the advantage of this approach is that a single unitary transformation acting on the rotor can simultaneously rotate and entangle many qubits. Limitations will, however, inevitably appear with the physical implementation. Furthermore, an encoding is practicable only if it can tolerate small errors affecting the physical system. Therefore, we use the stabilizer formalism \cite{gottesman98,aharonov08} to construct qubits that are robust against small errors in angular position and momentum. We then investigate practical approximations of our scheme and finally suggest a few implementations.

In Sec.~\ref{rotor} we briefly describe a quantum mechanical rotor and a qubit. We present in Sec.~\ref{encoding} a simple encoding to embed many qubits in a single rotor. We also give the corresponding universal set of quantum gates. In Sec.~\ref{protection} we provide a more general quantum error-correcting code to protect many qubits against small errors in angular position and momentum. It turns out that the protected qubits require an unlimited amount of energy. Thus we consider realistic approximations of the unphysical encoded qubits in Sec.~\ref{approximation}. We then suggest in Sec.~\ref{implementation} possible implementations in quantum optics, atom optics, and molecular physics. We briefly discuss the qudit case in Sec.~\ref{discussion}. We finally conclude in Sec.~\ref{conclusion}.

\section{Descriptions of a rotor and a qubit}\label{rotor}
In quantum mechanics the angular momentum $L$ for rotation around a fixed axis is a self-adjoint operator with a discrete and infinite spectrum. As we will see, this discreteness represents a very natural basis to encode qubits. 

We label the eigenstates of the operator $L$ with an integer ${\ell=0,\pm1,\pm2,\cdots}$ as
\begin{equation}
L \ket{\ell}= \ell \ket{\ell}.
\end{equation}
In this angular momentum basis $\{\ket{\ell}\}$, the orthogonality and completeness relations take the simple form
\begin{equation}
\bra{\ell}\ell'\rangle=\delta_{\ell,\ell'} \; \; \; \textrm{and} \; \; \; \sum_{\ell=-\infty}^{\infty}|\ell\rangle \! \langle \ell|=\openone.
\end{equation}
We further define the Fourier transform basis $\{\ket{\theta}\}$ of $\{\ket{\ell}\}$ as
\begin{equation}\label{t}
\ket{\theta}=\sum_{\ell=-\infty}^{\infty}e^{-i\ell\theta}\ket{\ell}
\end{equation}
so that the two observables are complementary, that is,
\begin{equation}
\left<\theta|\ell\right>=e^{i\ell\theta}.
\end{equation}
The angular position basis $\{\ket{\theta}\}$ is continuous and \mbox{$2\pi$-periodic}, ${\ket{\theta+2\pi}=\ket{\theta}}$. Accordingly, we have
\begin{equation}
\bra{\theta}\theta'\rangle=2\pi\delta^{(2\pi)}(\theta-\theta') \; \; \; \textrm{and} \; \; \; \int_{(2\pi)} \frac{d\theta}{2\pi} |\theta\rangle\!\langle\theta|=\openone,
\end{equation}
where ${\delta^{(2\pi)}(\phi)}$ is the \mbox{$2\pi$-periodic} delta function defined as ${\sum_{k=-\infty}^{\infty}\delta(\phi-2\pi k)}$ and the integration is over any ${2\pi}$ interval.

Let us now consider the operator $e^{i\alpha L}$, with real $\alpha$. Its action on a ket $\ket{\theta}$ reads 
\begin{equation}
e^{i\alpha L}\ket{\theta}=\ket{\theta-\alpha}.
\end{equation}
The operator $e^{i\alpha L}$ is therefore called the shift operator in angular position. It follows from the fact that $\ell$ is an integer that ${e^{2i\pi L}=\openone}$. We proceed to introduce $V$, the shift operator in angular momentum. It is defined through its action on $\{\ket{\ell}\}$ as
\begin{eqnarray}
V\ket{\ell}=\ket{\ell+1}.
\end{eqnarray}
Note for completeness that
\begin{eqnarray}
e^{i\alpha L}\ket{\ell}=e^{i\alpha \ell}\ket{\ell}\; \; \; \textrm{and} \; \; \;V\ket{\theta}=e^{i\theta}\ket{\theta}.
\end{eqnarray}
It is not too difficult to verify that these two shift operators satisfy the commutation relation
\begin{eqnarray}
e^{i\alpha L}V=Ve^{i\alpha L}e^{i\alpha}.
\end{eqnarray}
These two properties together with ${e^{2i\pi L}=\openone}$ define the so-called Weyl pair or Schwinger operators of a rotor. The notion of a Weyl pair is closely related to the Heisenberg-Weyl and generalized Pauli groups \cite{kibler08}.

A quantum mechanical system is entirely characterized by its Weyl pair. First, each Weyl pair fully defines the Hilbert space of the system. Second, the Weyl pair is algebraically complete, that is, any operator acting on the system can be expressed in terms of these two operators only. This makes the Weyl pair extremely useful for discussing the properties of a quantum mechanical system \cite{schwingerbook,bergesbook}.

As for the rotor we can introduce the Weyl pair ${(Z,X)}$ of a qubit. The operators $Z$ and $X$ are both unitary and hermitian and such that
\begin{eqnarray}\label{WeylPair}
Z^2=X^2=\openone \; \; \; \textrm{and} \; \; \;{ZX=-XZ}.
\end{eqnarray}
By construction, $Z$ and $X$ are Fourier transforms of each other and each one corresponds to the shift operator on the eigenbasis of the other.

Let us clarify here an important issue related to qubit encoding. For practical reasons, the 
involution property of the qubit Weyl pair might be withdrawn leading to the definition of a pseudo qubit. For example, one could only require $Z^2$ and $X^2$ to equal the projection operator onto the qubit subspace as seen in \cite{preskill01}. In this article, we focus our attention on genuine qubits for which the Hilbert space is the tensor product of the qubit space times the space for all other degrees of freedom.

\section{From a rotor to many qubits}\label{encoding}
For the sake of clarity we proceed in two steps. First, we show how to encode a single qubit on the basis of the angular momentum. This encoding can be schematically summarized as ${\textrm{rotor} = \textrm{qubit} \otimes \textrm{rotor}}$. Our mapping corresponds to a splitting of the angular momentum space into two isomorphic subspaces. One subspace corresponds to even angular momenta while the second subspace is associated with odd angular momenta. Such a decomposition is formally explained below but can be intuitively understood in terms of splitting the Fourier series of a \mbox{$2\pi$-periodic} function into even and odd parts. The splitting effectively provides a qubit while each partial \mbox{$2\pi$-periodic} Fourier series lead to a complete \mbox{$\pi$-periodic} Fourier series, that is, a new rotor. Second, we use the same procedure and this remarkable tensor product structure to further split the remaining rotor. We can repeatedly factorize the remaining rotor degree of freedom  and so create more and more qubit degrees of freedom. We shall now begin with a single qubit.

\subsection{Encoding a qubit in a rotor}\label{encodingsingle}
We want to encode a qubit subsystem in a rotor. Mathematically, we use the Weyl pair ${(e^{i\alpha L},V)}$ of the rotor to define the Weyl pair ${(Z,X)}$ of a qubit. Specifically, we choose the two unitary and hermitian operators $Z$  and $X$ \cite{englert06} as
\begin{eqnarray}\label{onequbit}
Z&=&e^{i \pi  L}=(-1)^{ L}, \nonumber \\
X&=&\frac{1}{2} \big( (\openone+Z)V^\dagger+V(\openone+Z) \big).
\end{eqnarray}
They obey the relation of~(\ref{WeylPair}) and hence they form the Weyl pair of a qubit. Note, furthermore, that the trace of $Z$ and the trace of $X$ vanish, as in the {$2$-dimensional} case. The action of the operator $Z$ is simple to understand as illustrated in Table~\ref{tab:table1}.
It only considers the parity of the eigenstate $\ket{\ell}$. The total Hilbert space is split into two eigenspaces, one with eigenvalue $+1$ corresponding to even angular momenta and one with eigenvalue $-1$ for the odd angular momenta.
Since the operator $Z$ is infinitely degenerate, at least another operator is necessary to obtain a Complete Set of Commuting Operators (CSCO) of the original rotor. Since two adjacent angular momenta are already discriminated by $Z$, the second operator of the CSCO can view them as identical. This can be done with the operator ${L_1=\lfloor L/2 \rfloor}$, where $\floor{x}$ denotes the floor function, that is, the largest integer that does not exceed $x$. This is shown in Table~\ref{tab:table1}.
\begin{table}
	\centering
		\begin{tabular}{cccrrrrrrrrrc}
		\hline\hline
			$L$ &\vline&$\dots$&$-4$&$-3$&$-2$&$-1$&~$0$&~$1$&~$2$&~$3$&~$4$&$\dots$\\ \hline
      $Z$ &\vline&$\dots$&$0$&$1$&$0$&$1$&$0$&$1$&$0$&$1$&$0$&$\dots$\\
      $L_1$ &\vline&$\dots$&$-2$&$-2$&$-1$&$-1$&$0$&$0$&$1$&$1$&$2$&$\dots$\\
      \hline\hline
     \end{tabular}
	\caption{A qubit and a residual rotor. In this simplest case, the encoding corresponds to the parity of the angular momentum eigenstate $\ket{\ell}$. The residual rotor is obtained by concatenating two adjacent angular momentum eigenstates.}
	\label{tab:table1}
\end{table}
Here, we attach the index $1$ to discriminate between the original rotor and what will be found to be a residual rotor. The two operators $Z$ and $L_1$ commute and one can write ${\ell=p+2 \floor{\ell/2}}$ where $\floor{\ell/2}$ is the eigenvalue of $L_1$ and $p$ such that ${z=(-1)^p}$ where $z$ is the eigenvalue of $Z$. Thus $Z$ and $L_1$ form a CSCO of the original rotor and completely specify the state of the system. We now have two alternative bases for the Hilbert space of the original rotor, either the eigenbasis $\{\ket{\ell}\}$ of the operator $L$ or the eigenbasis $\{\ket{p,\ell_1}\}$ of the CSCO ${(Z,L_1)}$. It remains to identify these two independent degrees of freedom and to make sure that their manipulation does not affect the other degree of freedom. In other words, first, we have to identify the two Weyl pairs of the two degrees of freedom and, second, not only $Z$ and $L_1$ have to commute but also their corresponding Weyl partners.

For the operator $Z$, the Weyl pair is that of a qubit. With respect to the operator $L_1$, one can substitute it with the operator $e^{i\alpha L_1}$ as they share the same eigenbasis. This suggests us to consider the pair ${\big(e^{i\alpha L_1},V_1\big)}$, where ${V_1=V^2}$. 
The commutation relation of these two unitary operators is
\begin{eqnarray}
e^{i\alpha L_1}V_1=V_1e^{i\alpha L_1} e^{i\alpha}
\end{eqnarray}
and ${e^{2i\pi L_1}=\openone}$. They form the Weyl pair of a rotor. As expected, this rotor is \mbox{$2\pi$-periodic}. Indeed one can introduce the eigenvalue $\theta_1$ of $V_1$ by  ${V_1\ket{p,\theta_1}= e^{i\theta_1}\ket{p,\theta_1}}$, and verify that
\begin{align}
\ket{0,\theta_1}&=\frac1{2}\Big(\ket{\theta}+\ket{\theta+\pi}\Big)=\ket{0,\theta_1+2\pi},\nonumber\\
\ket{1,\theta_1}&=\frac{e^{i\theta}}{2}\Big(\ket{\theta}-\ket{\theta+\pi}\Big)=\ket{1,\theta_1+2\pi},
\end{align}
where ${\theta=\theta_1/2}$. However, as noted earlier for the Fourier series of a \mbox{$2\pi$-periodic} function, the ket $\ket{\theta}$ can be decomposed as
\begin{align}
\ket{\theta}&=\ket{0,\theta_1}+e^{-i\theta_1/2}\ket{1,\theta_1},
\end{align}
where the kets $\ket{0,\theta_1}$ and $\ket{1,\theta_1}$ are \mbox{$\pi$-periodic} in $\theta$.

Finally one can compute the four commutators ${[Z,e^{i\alpha L_1}]}$, ${[Z,V_1]}$, ${[X,e^{i\alpha L_1}]}$, and ${[X,V_1]}$ and ascertain that they vanish. Consequently the rotor has been decomposed into a qubit subsystem and another rotor. It is crucial to notice the tensor product structure of this mapping which, as mentioned above, can be cast as ${\textrm{rotor} = \textrm{qubit} \otimes \textrm{rotor}}$. In mathematical terms this translates as ${{\cal H}\simeq\mathbb{C}^2\otimes {\cal H}}$, where $\cal{H}$ denotes the Hilbert space of the rotor. Here the qubit is encoded as a subsystem rather than as a subspace. The advantage of encoding in a subspace or in a subsystem may appear when considering a peculiar error model, a specific task to perform or, of course, practical considerations.

Now we can write the logical qubit basis states as
\begin{eqnarray}\label{basis1}
\ket{0,\psi_0}=\sum_{\ell \textrm{ even}}\alpha_{0\ell}\ket{\ell} \; \; \textrm{and} \; \;
\ket{1,\psi_1}=\sum_{\ell \textrm{ odd}}\alpha_{1\ell}\ket{\ell},
\end{eqnarray}
where the coefficient $\alpha_{0l}$ and $\alpha_{1l}$ are complex numbers such that ${\sum_{\ell \textrm{ even}}|\alpha_{0\ell}|^2=\sum_{\ell \textrm{ odd}}|\alpha_{1\ell}|^2=1}$ but otherwise unrestricted. The ket is labeled with the parity of the eigenvalue of the $Z$ operator while $\psi_0$ and  $\psi_1$ specify the state of the residual rotor, here assumed to be in a pure state. There is a freedom in the choice of the residual rotor states or equivalently in the coefficients $\alpha_{i\ell}$, ${i=0,1}$. This is a consequence of the infinitely degenerate spectrum of the operator $Z$. We will use this freedom to our advantage. More specifically we will exploit the degeneracy in two ways: first to encode many qubits, second, to protect these qubits against errors in the angular position and momentum. The encoding of many qubits is investigated in the remainder of this section while their protection against errors in realistic realizations will be thoroughly studied in Sec.~\ref{protection}.

\subsection{Encoding many qubits in a rotor}

A consequence of the tensor product structure is to allow the encoding of many more logical qubits. Indeed nothing prevents us from exploiting the same encoding on the residual rotor to obtain two qubits, then three and even more. It should be clear that the above encoding allows in principle to encode as many qubits as desired. The limitation will of course come from the physical implementation as the more qubits we encode, the larger the accessible angular momenta must be.

We shall now pay due attention to the Weyl pairs of different qubits. First we add an index $j$, ${j=1, \cdots, N}$, to the operators $Z$ and $X$ to specify which qubit is considered. The encoding of three qubits is exemplified in Table~\ref{tab:table2}.
\begin{table}
	\centering
		\begin{tabular}{cccrrrrrrrrrc}
		\hline\hline
			$L$ &\vline&$\dots$&$-4$&$-3$&$-2$&$-1$&~$0$&~$1$&~$2$&~$3$&~$4$&$\dots$\\ \hline
      $Z_1$ &\vline&$\dots$&$0$&$1$&$0$&$1$&$0$&$1$&$0$&$1$&$0$&$\dots$\\
      $Z_2$ &\vline&$\dots$&$0$&$0$&$1$&$1$&$0$&$0$&$1$&$1$&$0$&$\dots$\\
      $Z_3$ &\vline&$\dots$&$1$&$1$&$1$&$1$&$0$&$0$&$0$&$0$&$1$&$\dots$\\
      $L_3$ &\vline&$\dots$&$-1$&$-1$&$-1$&$-1$&$0$&$0$&$0$&$0$&$0$&$\dots$\\
      \hline\hline
     \end{tabular}
	\caption{Encoding of three qubits in a rotor. Here three qubits are encoded by repeatedly factorizing the rotor degree of freedom. The residual rotor is obtained by concatenating $2^3$ adjacent angular momentum eigenstates.}
	\label{tab:table2}
\end{table}
This new mapping can be compared to the standard binary mapping recalled in Table~\ref{tab:table3}. For the binary mapping, the encoding of the first qubit is anti-symmetric with respect to ${\ell=0}$ while the encoding of all the remaining qubits is symmetric with respect to ${\ell=0}$. This is in strong contrast with the new proposed mapping which is invariant under translation of $2^n$ angular momenta, where $n$ is the number of qubits. For example, the encoding presented in Table~\ref{tab:table2} in invariant under translation of $8$ angular momenta.
\begin{table}
	\centering
		\begin{tabular}{cccrrrrrrrrrc}
		\hline\hline
			$L$ &\vline&$\dots$&$-4$&$-3$&$-2$&$-1$&~$0$&~$1$&~$2$&~$3$&~$4$&$\dots$\\ \hline
      $Z_1$ &\vline&$\dots$&$1$&$1$&$1$&$1$&$0$&$0$&$0$&$0$&$0$&$\dots$\\
      $Z_2$ &\vline&$\dots$&$0$&$1$&$0$&$1$&$0$&$1$&$0$&$1$&$0$&$\dots$\\
      $Z_3$ &\vline&$\dots$&$0$&$1$&$1$&$0$&$0$&$0$&$1$&$1$&$0$&$\dots$\\
      \hline\hline
     \end{tabular}
	\caption{Binary encoding of three qubits. This encoding differs from the encoding presented in Table~\ref{tab:table2}.}
	\label{tab:table3}
\end{table}
By construction we can write the respective Weyl operators $Z_j$ and $X_j$ for the $j$th qubit as
\begin{eqnarray}\label{firstencoding}
Z_j&=&(-1)^{\floor{ L/2^{j-1}}}, \nonumber \\
X_j&=&\frac{1}{2} \big( (\openone+Z_j)V^{\dagger 2^{j-1}}+V^{2^{j-1}}(\openone+Z_j) \big).
\end{eqnarray}
The total Hilbert space is here again divided into two eigenspaces, corresponding to even and odd values of ${\floor{{\ell}/2^{j-1}}}$, respectively. Evidently, the single qubit in Sec.~\ref{encodingsingle} corresponds to the first qubit ${j=1}$. Furthermore, the Weyl pair of the residual rotor after encoding $N$ qubits is given by ${\big(e^{i\alpha\floor{L/2^N}} ,V^{2^N}\big)}$.
To illustrate the mapping presented above, let us explicitly write down the computational basis of the two-qubit subspace:
\begin{eqnarray}\label{basis2}
\ket{00,\psi_{00}}&=&\sum_{\ell=-\infty}^{\infty}\alpha_{00\ell}\ket{4\ell}, \nonumber \\ \ket{01,\psi_{01}}&=&\sum_{\ell=-\infty}^{\infty}\alpha_{01\ell}\ket{1+4\ell}, \nonumber \\
\ket{10,\psi_{10}}&=&\sum_{\ell=-\infty}^{\infty}\alpha_{10\ell}\ket{2+4\ell}, \nonumber \\ \ket{11,\psi_{11}}&=&\sum_{\ell=-\infty}^{\infty}\alpha_{11\ell}\ket{3+4\ell},  
\end{eqnarray}
with the proper normalization ${\sum_{\ell=-\infty}^{\infty}|\alpha_{jk\ell}|^2=1}$, for ${j,k=0,1}$. As already mentioned, the freedom in the residual rotor states will be used to ensure protection against shifts in angular position and momentum.

We finally complete the set of single qubit operations, which can be constructed from $Z_j$ and $X_j$, with an entangling two-qubit gate to obtain a universal set of quantum gates. We choose the phase-shift gate $R$ as the entangling two-qubit operation. This unitary transformation can be simply expressed in the $\{\ket{\ell}\}$ basis as
\begin{eqnarray}
R_{jk}&=&\sum_{\ell=-\infty}^{\infty} \Big(\frac{1+(-1)^{\ell_j}}{2} + \frac{1-(-1)^{\ell_j}}{2}(-1)^{\ell_k}\Big) |\ell \rangle \langle \ell|\nonumber \\
&=& R_{kj},
\end{eqnarray}
where the indexes $j$ and $k$ denote the two qubits concerned by the gate so that ${\ell_j=\floor{\ell/2^{j-1}}}$ and ${\ell_k=\floor{\ell/2^{k-1}}}$.

\section{Protection of many qubits}\label{protection}
The qubits encoded in this manner are vulnerable to errors. In the case of a single qubit, the logical qubits $\ket{0}$ and $\ket{1}$ are encoded in even angular momenta and odd angular momenta, respectively. Therefore, a state $\ket{0}$ affected by a unit shift ${\ell \rightarrow \ell+1}$ will not be distinguishable from an unaffected state $\ket{1}$, as can be seen for Eq.~(\ref{basis1}). Distinguishability between $\ket{0}$ and $\ket{1}$ is then lost and it is thus necessary to provide an encoding which protects the logical qubits against small errors.

In a very enlightening presentation \cite{preskill01}, Gottesman {\it et al.} use the stabilizer formalism to protect a qudit (subspace) embedded in a infinite-dimensional space. Here we exploit the same technique to protect not only one but many qubits (subsystems). Actually we will show in Sec.~\ref{discussion} that our results also apply to qudits. The technique used in \cite{preskill01} slightly differs from the common use of the stabilizer formalism where $k$ logical qubits are encoded in ${K \geq k}$ physical qubits \cite{shor95,gottesman98}. But the basic ideas are the same.

In a nutshell the stabilizer formalism works as follows. We consider a set of orthogonal states that we want to protect against errors and call this subspace the code space. Within this code space, each state or {\it code word} is identified by the eigenvalue of a suitable operator. Errors are represented by unitary transformations that move the code space to another subspace. The different subspaces corresponding to different errors have to be distinguishable to allow correction. Therefore, they must be orthogonal. A simple solution to impose this orthogonality is to associate each subspace with a different eigenvalue of a suitably chosen set of unitary transformations. Eigenspaces are orthogonal and can therefore be perfectly distinguished. This chosen set of unitary transformations is the so-called stabilizer.  By convention we identify the code space as the $+1$ eigenspace of the stabilizer. The errors will then move the code words from the $+1$ eigenspace to another, therefore orthogonal, eigenspace of the stabilizer. In this ideal case the error is identified by reading out the eigenvalues of the stabilizer. Correction is then performed by applying the inverse operator of the identified error. The $+1$ eigenspace is stabilized and contains our protected states. 

Here we are interested in two types of errors: continuous drifts for the angular position and discrete shifts for the angular momentum. They can be written in terms of the shift operators as
\begin{eqnarray}
E_{\theta}(\epsilon)=e^{i \epsilon  L} \; \; \; \textrm{and} \; \; \;  E_{L}({\rm e})=V^{{\rm e}},
\end{eqnarray}
where $\epsilon$ is real and ${\rm e}$ is an integer. To detect these two types of errors, two commuting unitary transformations are required. They also take the form of shift operators. A \mbox{$n$-dimensional} subspace of a rotor can be protected against shifts in angular momentum and angular position using the stabilizer \cite{preskill01}
\begin{eqnarray}
S_{\theta}=V^{m} \; \; \; \textrm{and} \; \; \;  S_{L}=e^{2 i \pi \frac{n}{m} L},
\end{eqnarray}
where $m$ is a free parameter related to the maximum amount of correctable errors. $S_{\theta}$ will be used to read off the amplitude $\epsilon$ of the error in $\theta$, while $S_{L}$ will be used to read off the amplitude ${\rm e}$ of the error in $L$. The above stabilizer can correct drifts in angular position up to ${|\epsilon| < \pi/m}$ and shifts in momentum up to ${| {\rm e} |<m/(2n)}$. For simplicity we now define ${\Delta \theta>0}$ and ${\Delta L>0}$ as the maximal correctable error in $\theta$ and $L$. For a fixed $n$, an increase in $m$ will lead to an increase of ${\Delta L}$ but a decrease of ${\Delta \theta }$. A trade-off follows, depending on which type of error is more critical for a given implementation or application.
The stabilized space is composed of $n$ code words or protected states 
\begin{eqnarray}\label{codeword}
\ket{k[n],1,1}=\sqrt{m}\sum_{\ell=-\infty}^{\infty} |m(k/n+\ell)\rangle,
\end{eqnarray}
where $k$ is an integer. Note that the factor $\sqrt{m}$ is for convenience. We use the label $k[n]$ to emphasize that only $k$ modulo $n$ is relevant as can be seen from the above definition. The label ${1,1}$ refers to the stabilizer's eigenvalues. Let us also note that ${m/n}$ has to be an integer to ensure the existence of a \mbox{$n$-dimensional} stabilized space. Otherwise only the state ${\ket{0[n],1,1}=\sqrt{m}\sum_{\ell=-\infty}^{\infty} |m\ell\rangle}$ among the $n$ states $\ket{k[n],1,1}$ is well-defined. We denote this integer by $r$. In the angular momentum basis the code words are equally weighted superpositions of shifted states. The code words take the form of a finite sum in the angular position basis 
\begin{eqnarray}\label{solangularposition}
\ket{k[n],1,1}=\frac{1}{\sqrt{m}} \sum_{j=0}^{m-1} e^{-2 i \pi kj/n} \ket{\theta=\frac{2 \pi}{m}j},
\end{eqnarray}
where the phases are powers of the $n$th root of unity.

Since we want to protect $N$ qubits, we have to identify the \mbox{$n$-dimensional} protected subsystem with the space spanned by the $N$ qubits. Technically, not only the dimensions have to match, i.e.,\ ${n=2^N}$, but also the qubit operators $Z_j$ and $X_j$ have to commute with the other qubits' operators ($Z_k$ and $X_k$, ${k \neq j}$) as well as with the stabilizer. A suitable encoding is provided by the following pairs
\begin{eqnarray}\label{firstencodingbis}
Z_j&=&(-1)^{\floor{ L/(2^{j-1}r)}}, \nonumber \\
X_j&=&\frac{1}{2} \big( (\openone+Z_j)V^{\dagger 2^{j-1}r}+V^{2^{j-1}r}(\openone+Z_j) \big).
\end{eqnarray}
The CSCO of the original rotor now is ${(Z_1,\cdots,Z_N,S_L,S_{\theta})}$. To prove it, let us consider an alternative writing of the basis $\{\ket{\ell}\}$.

The eigenvalues of $Z_j$ are of the form $(-1)^{p_j}$ where ${p_j=0,1}$. The eigenvalues of $S_L$ are of the form $e^{2i\pi q/r}$ where ${q=0,\cdots,r-1}$ while the eigenvalues of ${\lfloor L/m\rfloor}$ are $\floor{\ell/m}$. The identity
\begin{eqnarray}
\ell=\sum_{j=1}^N p_j 2^{j-1}r +q + \floor{\ell/m} m,
\end{eqnarray}
therefore, provides a relabeling of the kets $\ket{\ell}$ in terms of the quantum numbers $p_j$, $q$, and $\floor{\ell/m}$. It follows that the operators ${(Z_1,\cdots,Z_N,S_L,S_{\theta})}$ commute and form a CSCO of the original rotor. Since $V^m$ and ${\lfloor L/m\rfloor}$ span the same subspace, the set ${(Z_1,\cdots,Z_N,S_L,S_{\theta})}$ is a CSCO of the rotor too.

An example of such CSCO is illustrated in Table~\ref{tab:table4} for two qubits and ${m=6}$.
\begin{table}
       \centering
               \begin{tabular}{cccrrrrrrrrrccccc}
               \hline\hline
                       $L$ &\vline&$\dots$&~$0$&~$1$&~$2$&~$3$&~$4$&~$5$&~$6$&~$7$&~$8$&~$9$&$10$&$11$&$12$&$\dots$\\ \hline
     $Z_1$ &\vline&$\dots$&$0$&$0$&$0$&$1$&$1$&$1$&$0$&$0$&$0$&$1$&$1$&$1$&$0$&$\dots$\\
     $Z_2$ &\vline&$\dots$&$0$&$0$&$0$&$0$&$0$&$0$&$1$&$1$&$1$&$1$&$1$&$1$&$0$&$\dots$\\
     $S_L$ &\vline&$\dots$&$0$&$1$&$2$&$0$&$1$&$2$&$0$&$1$&$2$&$0$&$1$&$2$&$0$&$\dots$\\
     $\floor{L/m}$ &\vline&$\dots$&$0$&$0$&$0$&$0$&$0$&$0$&$0$&$0$&$0$&$0$&$0$&$0$&$1$&$\dots$\\
     \hline\hline
    \end{tabular}
       \caption{CSCO for two qubits and ${m=6}$. Here the quantum numbers $p_1$, $p_2$, $q$, and $\floor{\ell/m}$ of the operators $Z_1$, $Z_2$, $S_L$, and $\floor{L/m}$ uniquely specify the angular momentum eigenstate $\ket{\ell}$.}
       \label{tab:table4}
\end{table}
For an easier reading of this table, the eigenvalues of the operators $Z_j$ and  $S_L$ are labeled by the powers $p_j$ and $q$ instead of the eigenvalues themselves. For instance the eigenvalues of $S_L$ for two qubits and ${m=6}$ are ${1,e^{2i\pi/3},e^{-2i\pi/3}}$ but we chose to label them as ${0,1,2}$.

One must not fail to note that ${\big(e^{i\alpha\floor{L/m}},V^m\big)}$ constitutes the Weyl pair of a rotor so that the last degree of freedom corresponds to a residual rotor. Moreover $X_j$ commute with $e^{i\alpha\floor{L/m}}$ and $V^m$, with the other qubits' Weyl pairs as well as with $S_L$, so that we can finally write
\begin{eqnarray}
\textrm{rotor} = \textrm{qubits} \otimes (\textrm{error in $\ell$}) \otimes \textrm{rotor}.
\end{eqnarray}
As a result we can still encode more qubits from the residual rotor and the situation is similar to that of Sec.~\ref{encoding}. In principle we could also look at the Weyl pair associated with $S_L$ and name the corresponding degree of freedom however this is irrelevant to the present study.

The encoding of Eq.~(\ref{firstencodingbis}) allows error correction in angular momentum up to ${\Delta L<r/2}$. Thus we can choose to define $r$ from the maximal number of correctable errors, that is, ${r=2\Delta L+1}$. An even $r$ is also conceivable however the resulting encoding would be less compact and will therefore require to access greater angular momenta. The simple encoding provided in Sec.~\ref{encoding} is just a special case where ${\Delta L=0}$ (${r=1}$) and would correspond to the stabilizer element ${S_L=\openone}$ (${m=n}$). The general protected qubits are then given by
\begin{align}
\ket{k[n],1,1}\!=\!\sqrt{2^N (2 \Delta L +1)}\!\sum_{\ell=-\infty}^{\infty}\!\ket{(2 \Delta L +1)(k+2^N\ell)}\!.
\end{align}
Since the protected states are $N$ protected qubits, it is useful to label them with the eigenvalues of the operators $Z_j$. In the following we drop the eigenvalues of the stabilizer to lighten the notation. Let us illustrate what we have learned with three examples. First we can consider a single qubit with no protection against shift in the angular momentum, that is, $N=1$, ${\Delta L=0}$, ${\Delta \theta < \pi/2}$. The computational basis is then given by
\begin{eqnarray}\label{protectedqubit1}
\ket{0}&=&\sqrt{2} \sum_{\ell=-\infty}^{\infty}\ket{2\ell},\nonumber \\
\ket{1}&=&\sqrt{2} \sum_{\ell=-\infty}^{\infty}\ket{1+2\ell}.
\end{eqnarray}
Equivalently, these states can be written in the angular position basis as:
\begin{eqnarray}\label{protectedqubit2}
\ket{0}&=&\frac{1}{\sqrt{2}}(\ket{0}+\ket{\pi}), \nonumber \\
\ket{1}&=&\frac{1}{\sqrt{2}}(\ket{0}-\ket{\pi}).
\end{eqnarray}
For a single qubit with protection against a unit shift in angular momentum (${\Delta L=1}$ and ${\Delta \theta < \pi/6}$), we obtain the two protected states
\begin{eqnarray}
\ket{0}&=&\sqrt{6} \sum_{\ell=-\infty}^{\infty}\ket{6\ell}, \nonumber \\
\ket{1}&=&\sqrt{6} \sum_{\ell=-\infty}^{\infty}\ket{3+6\ell}.
\end{eqnarray}
A more interesting instance is concerned with two qubits. If we ask for protection against unit shifts in angular momentum (${\Delta L=1}$ and ${\Delta \theta < \pi/12}$), we end up with
\begin{eqnarray}
\ket{00}&=&2\sqrt{3}\sum_{\ell=-\infty}^{\infty}\ket{12 \ell}, \nonumber \\
\ket{10}&=&2\sqrt{3}\sum_{\ell=-\infty}^{\infty}\ket{3+12\ell}, \nonumber \\
\ket{01}&=&2\sqrt{3}\sum_{\ell=-\infty}^{\infty}\ket{6+12\ell}, \nonumber \\
\ket{11}&=&2\sqrt{3}\sum_{\ell=-\infty}^{\infty}\ket{9+12\ell}.
\end{eqnarray}

As already mentioned earlier, these ideal code words are unphysical. This can be seen for example in Eq.~(\ref{protectedqubit1}) and Eq.~(\ref{protectedqubit2}). These states take the form of an infinite sum of equally weighted angular momenta and are perfectly localized in angular position. In the following section we will investigate physical approximations of the ideal protected qubits.

\section{Physical approximations}\label{approximation}
The code words, and in particular the protected qubits, are infinitely squeezed states in angular position. With implementation in mind we should investigate the behavior of realistic approximations of the protected qubits. These are the physical approximations. Clearly, finite squeezing will inevitably lead to additional errors. Sometimes these errors are within the range of correctable errors, sometimes they are not. We calculate the probability of noncorrectable errors for physical approximations of the protected qubits.

We consider four physical approximations. Two approximations pattern a finite squeezing in position. The first one relies on a truncated Gaussian wave function \cite{padgett04} while the second is a power of cosine wave function. The remaining two approximations are better envisioned in the angular momentum basis. One approximation resorts on a Gaussian envelop for the infinite sum of angular momenta whereas the last instance focuses on a finite sum which we name grating. We study analytically the truncated Gaussian approximation and provide numerical plots for the three other approximations.

\subsection{Error probability}
We want to define the probability of noncorrectable errors. For simplicity we first restrict ourselves to a single qubit only protected against drifts in angular position (${\Delta L =0}$ and ${\Delta\theta=\pi/2}$). But the generality of the arguments remains and the formulas derived below are immediately generalized to larger $N$ and ${\Delta L}$. In view of Eq.~(\ref{protectedqubit2}), the errors are specified on the two basis states $\ket{0}$ and $\ket{\pi}$. Moreover measurements are here performed modulo $\pi$ in order to avoid destroying the superposition. Since only an error smaller than ${\Delta\theta}$ can be corrected while the approximated basis state spreads all over the range ${[0,2\pi]}$, some noncorrectable errors happen. For the state $\ket{0}$, any value of the angular position measured in the range ${[-\pi/2,\pi/2]}$ can be corrected while values within ${[-\pi,-\pi/2]}$ and ${[\pi/2,\pi]}$ cannot. For the state $\ket{\pi}$, the situation is opposite. Any value of the angular position measured within ${[-\pi,-\pi/2]}$ and ${[\pi/2,\pi]}$ can be corrected while values in ${[-\pi/2,\pi/2]}$ cannot. This naturally leads to the classical probability of error for the approximated states \mbox{$\ket{0}$ and $\ket{\pi}$}:
\begin{eqnarray}
p_e(0)&=&\int_{-\pi}^{-\frac{\pi}{2}}\textrm{d}\theta \; |\langle\theta\ket{0}|^2+\int_{\frac{\pi}{2}}^{\pi}\textrm{d}\theta \; |\langle\theta\ket{0}|^2, \nonumber \\
p_e(\pi)&=&\int_{-\frac{\pi}{2}}^{\frac{\pi}{2}}\textrm{d}\theta \; |\langle\theta\ket{\pi}|^2.
\end{eqnarray}
In these equations and in the succeeding ones, unless stated otherwise, the symbols $\ket{0}$ and $\ket{\pi}$ refer to the approximation of the basis states.
Taking into account the symmetry of the wave function, we end up with
\begin{eqnarray}
p_e(0)=p_e(\pi)=2\int_{0}^{\frac{\pi}{2}}\textrm{d}\theta \; |\langle\theta\ket{\pi}|^2.
\end{eqnarray}

We can now define the probability of error for any state diagonal in the basis ${\{\ket{0},\ket{\pi}\}}$. For a given density matrix ${\rho=\alpha_0 |0\rangle \! \langle0|+\alpha_{\pi} |\pi\rangle \! \langle\pi|}$ with ${\alpha_0+\alpha_{\pi}=1}$, the error probability is 
\begin{eqnarray}\label{error}
p_e(\rho)=\alpha_0 p_e(0)+ \alpha_{\pi} p_e(\pi).
\end{eqnarray}
This immediately leads to
\begin{eqnarray}\label{errorproba}
p_e(\rho)=2\int_{0}^{\frac{\pi}{2}}\textrm{d}\theta \; |\langle\theta\ket{\pi}|^2.
\end{eqnarray}

Any equally-weighted classical mixture of the logical states ${\ket{0}=(\ket{0}+\ket{\pi})/\sqrt{2}}$ and ${\ket{1}=(\ket{0}-\ket{\pi})/\sqrt{2}}$ is diagonal in ${\{\ket{0},\ket{\pi}\}}$. Indeed, the off-diagonal elements nicely vanish.

However, we cannot apply the above definition to more general mixtures ($\alpha_0$ and $\alpha_1$ different from one half) where the off-diagonal terms 
${|0\rangle \! \langle\pi|}$ and ${|\pi\rangle \! \langle0|}$ do not vanish. Nevertheless since the error is defined in the $\{\ket{\theta}\}$ basis we can reasonably extend the definition of the error probability to states diagonal in the $\{\ket{\theta}\}$ basis.

Any good approximate state should have a sharp peak and fast decreasing tails. Therefore, a well approximated state should have nearly vanishing cross terms ${\langle \theta\ket{0} \! \bra{\pi} \theta \rangle}$ and ${\langle \theta\ket{\pi} \! \bra{0} \theta \rangle}$ leading to an almost diagonal state in the $\{\ket{\theta}\}$ basis. Under this constraint, the error probability for any approximated states is still given by Eq.~(\ref{errorproba}).

The generalization to any number of qubits and any ${\Delta L}$ is immediately given by
\begin{eqnarray}\label{sol}
p_e(\rho)=2\int_{0}^{\pi(1-\frac{1}{m})}\textrm{d}\theta \; |\langle\theta\ket{\pi}|^2.
\end{eqnarray}
In the general case, the measurement of the angular position is modulo ${2\pi/m}$ to avoid destroying the superposition of Eq.~(\ref{solangularposition}).
We compute the probability of error for the so-called truncated Gaussian states and numerically evaluate it for the three other approximations.

\subsection{Truncated Gaussian states}
The wave function of a truncated Gaussian state centered at $\theta_0$ and with a degree of squeezing $\xi$ is given by
\begin{eqnarray}\label{trunc}
\Psi_{\xi}(\theta-\theta_0)&=&\frac{\xi \sqrt{2 \pi} }{\sqrt{C_\xi}} e^{-\frac{\xi^2}{2}(\theta-\theta_0)^2},
\end{eqnarray}
where $\theta_0$ is in ${[0,2\pi]}$ and ${\theta_0 - \pi< \theta < \theta_0 + \pi}$. The normalization is ${C_{\xi}=\xi \sqrt{\pi}  \; \textrm{erf}(\pi \xi)}$  where
\begin{eqnarray}
\textrm{erf}(x)=\frac2{\sqrt{\pi}} \int_0^x \textrm{dt} \; e^{-t^2}
\end{eqnarray}
denotes the error function. As required, in the limit of infinite squeezing the non-normalized wave function tends to ${2\pi}$ times the delta function. Note that formally the truncated Gaussian function is not \mbox{$2\pi$-periodic} unlike the braket ${\langle \theta | \theta_0 \rangle=2\pi\delta^{(2\pi)}(\theta-\theta_0)}$ however we only consider this function in the range ${\theta_0 - \pi< \theta < \theta_0 + \pi}$. In the limit of no squeezing the truncated Gaussian wave function reduces to the unit constant. Moreover the overlap ${\langle \theta\ket{0} \! \bra{\pi} \theta \rangle}$ decreases as $\xi$ increases. Together with Eq.~(\ref{sol}) these properties imply that the error probability vanishes in the limit of infinite squeezing while in the limit of no squeezing, the error probability tends to the probability of a pure guess, that is, ${1-1/(2^N(2\Delta L +1))}$.

\begin{figure}[ht]
\centering
\includegraphics[scale=.6]{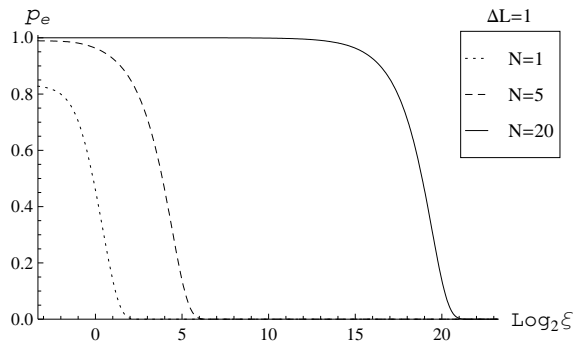}
\caption{Error probability for the truncated Gaussian in the case of $1$, $5$, and $20$ encoded qubits and protection against a single shift in angular momentum. When the physical parameter $\xi$ is much larger than the dimension of the qubits subspace $2^N$, the error probability decays faster than a Gaussian. When $\xi$ and $2^N$ are of the same order, the error probability is close to $10^{-5}$.}
\label{fig:plottruncgauss}
\end{figure}

Using Eqs.~(\ref{sol})-(\ref{trunc}), the probability of error for the truncated Gaussian reads 
\begin{eqnarray}
p_e^{1 \textrm{qubit}}=1-\frac{\textrm{erf}(\pi \xi/2)}{\textrm{erf}(\pi \xi)}
\end{eqnarray}
for a single qubit. More generally, the error probability for $N$ qubits is
\begin{eqnarray}
p_e^{\textrm{N qubits}}=1-\frac{\textrm{erf}\Big(\pi \xi / 2^N(2\Delta L +1)\Big)}{\textrm{erf}(\pi \xi)}.
\end{eqnarray}
When $\xi$ is of the order of ${m=2^N}$, the error probability ${p_e^{\textrm{N qubits}}}$ is close to $10^{-5}$ for arbitrary $N$. In the limit of large squeezing, i.e.,\ ${\xi \gg m}$, $p_e^{\textrm{N qubits}}$ scales as ${m e^{-(\frac{\pi \xi}{m})^2}/(\pi^{3/2}\xi)}$.
The error probability is plotted in terms of the degree of squeezing $\xi$ in Fig.~\ref{fig:plottruncgauss} for one, five, and twenty qubits and ${\Delta L=1}$.

\subsection{Other approximations}

Here we consider the three remaining approximations. We only give their definitions and plot their error probabilities in terms of the relevant physical parameter for one and five qubits and ${\Delta L=1}$.

\begin{figure}[ht]
\centering
\includegraphics[scale=.6]{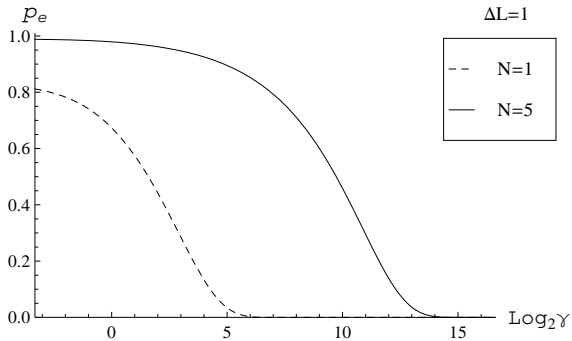}
\caption{Error probability for the power of cosine in the case of $1$ and $5$ encoded qubits and protection against a single shift in angular momentum. When the physical parameter $\gamma$ and the dimension of the qubits subspace $2^N$ are of the same order, the error probability is close to $1$. A reasonable approximation therefore requires ${\gamma \gg 2^N}$.}
\label{fig:plotcos}
\end{figure}

First the power of cosine wave function is defined as
\begin{eqnarray}
\Psi_{\gamma}(\theta-\theta_0)=\frac{2 \pi}{\sqrt{C_{\gamma}}}\Big(\textrm{cos}\Big(\frac{\theta-\theta_0}{2}\Big)\Big)^{\gamma},
\end{eqnarray}
with the suitable normalization $C_{\gamma}$. Here $p_e^{\textrm{N qubits}}$ is close to $1$ when  $\gamma$ and $m$ are of the same order. Therefore, the power $\gamma$ should be much larger than $m$ to keep the level of error reasonably low. The graphs of the error probability in terms of the power $\gamma$ for one and five qubits are shown in Fig.~\ref{fig:plotcos}.

\begin{figure}[ht]
\centering
\includegraphics[scale=.6]{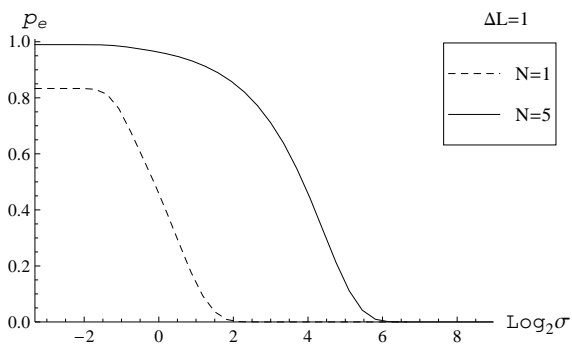}
\caption{Error probability for the Gaussian envelop in the case of $1$ and $5$ encoded qubits and protection against a single shift in angular momentum. As for the truncated Gaussian, the error probability is close to $10^{-5}$ when the physical parameter $\sigma$ and the dimension of the qubits subspace $2^N$ are of the same order.}
\label{fig:plotgaussenv}
\end{figure}

\begin{figure}[ht]
\centering
\includegraphics[scale=.6]{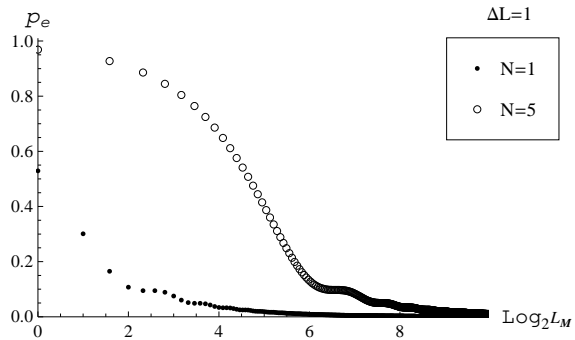}
\caption{Error probability for the grating in the case of $1$ and $5$ encoded qubits and protection against a single shift in angular momentum. For a number of slits $L_M$ of the order of the dimension of the qubits subspace $2^N$, the error probability is close to $10^{-1}$. Consequently, the number of slits must be much larger than $2^N$ to obtain a satisfactory approximation.}
\label{fig:plotgrating}
\end{figure}

Second the error probability for the Gaussian envelop
\begin{eqnarray}
\Psi_{\sigma}(\theta-\theta_0)=\frac{1}{\sqrt{C_{\sigma}}} \sum_{\ell=- \infty}^{\infty} e^{-\ell^2/(2\sigma^2)} e^{i \ell (\theta-\theta_0)},
\end{eqnarray}
where $C_{\sigma}$ is the required normalization, is plotted in terms of the width $\sigma$ in Fig.~\ref{fig:plotgaussenv}.
Similarly to the truncated Gaussian case, the error probability for the Gaussian envelop is close to $10^{-5}$ for $\sigma$ and $m$ of the same order.
Finally, the error probability for the grating
\begin{eqnarray}
\Psi_{L_M}(\theta-\theta_0)=\frac{1}{\sqrt{2L_M+1}} \sum_{\ell=-L_M}^{L_M} e^{i \ell (\theta-\theta_0)}
\end{eqnarray}
is represented in terms of the number of slits $L_M$ in Fig.~\ref{fig:plotgrating}.
Note that the range of this plot starts at $0$ (i.e.,\ ${L_M=1}$) and the spacing between two points is not constant since $L_M$ has to be an integer. The error probability $p_e^{\textrm{N qubits}}$ tends to $10^{-1}$ for when  $L_M$ and $m$ are of the same order. Consequently, the number of slits must be much larger than $m$ to obtain a low probability of error.

\section{Implementation}\label{implementation}
We discuss a possible implementation of a quantum rotor in quantum optics and briefly mention two other realizations using atom optics and molecular physics.

In quantum optics, the orbital angular momentum of light can be used as a quantum rotor. It has been shown that any  light beam  with  an  amplitude distribution of the form ${u(r,\theta,z)=u_0(r,z)e^{i\ell\theta}}$, where $\ell$ is an integer, carries a quantized orbital angular momentum around the beam axis $z$ \cite{Allen92}. Physically the distribution ${u(r,\theta,z)=u_0(r,z)e^{i\ell\theta}}$ can be obtained using a Laguerre--Gaussian (LG) light beam. There, each photon carries an angular momentum ${\ell\hbar}$. In this context, the angular momentum is also called LG mode. With today's technology LG modes can be easily produced and manipulated \cite{Leach02,Wei03}. For the proposed encoding, an infinite superposition of LG modes is required. Such a superposition can be experimentally realized, e.g., using a single LG mode state which propagation axis is suitably displaced \cite{Molina02}. Obviously, in any experimental scheme only a superposition of a finite number of modes is achieved and approximations, like the grating or the Gaussian envelop presented in the Sec.~\ref{approximation}, have to be taken into account. Any operation on the encoded qubits can then be realized with holograms, lenses, and linear optical elements and the measurements are carried out with well-established interferometric technologies.

\begin{figure}[ht]
\centering
\includegraphics[scale=.7]{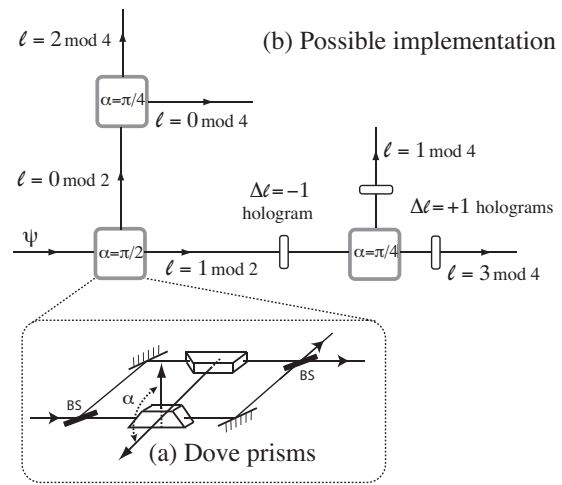}
\caption{Possible experimental setup. (a) A Mach--Zehnder interferometer with two Dove prisms at each arm is used to sort angular momentum states. By fixing the relative angle between the prisms to ${\alpha=\pi/2}$ one can sort even and odd angular momenta, as described in \cite{Leach02}. (b) To encode two qubits with ${\Delta L=0}$, we sort the beam into four ports where each port corresponds to orbital angular momenta modulo $4$. The sorting is done with Mach--Zehnder interferometers with a Dove  prism inserted into each arm and holograms. Once the modes are sorted, desired operations on the first and the second encoded qubits can be simultaneously performed.}
\label{fig:ex}
\end{figure}

Let us first consider the encoding of a single qubit with ${\Delta L=0}$. In this simplest case, the states of the computational basis are superpositions of either even or odd modes. To address the even and odd modes we use the {\it sorting} technique introduced in \cite{Leach02} where a Mach--Zehnder interferometer with a Dove  prism inserted into each arm is used, as shown in Fig.~\ref{fig:ex}a. By fixing the relative angle between the two Dove prisms to ${\pi/2}$ and correctly adjusting the path length of the interferometer, one can ensure that even modes appear in one port while odd modes appear in the other port of the interferometer. The qubit operation, e.g., $Z$ is realized by applying a \mbox{$\pi$-phase} shift to the odd modes, while the qubit operation $X$ is realized by passing the even modes through a hologram which increases all the momenta by one unit, and passing the odd modes through a hologram which decreases all the momenta by one unit. Finally, after applying the desired operations, the two arms can be recombined.

To encode two qubits with ${\Delta L=0}$, we sort the beam into four ports, as illustrated in Fig.~\ref{fig:ex}b. The first port corresponds to orbital angular momenta $0$ modulo $4$, the second port corresponds to $1$ modulo $4$, and similarly for the third and fourth ports. This sorting is done by cascading Mach--Zehnder interferometers (with a Dove  prism inserted into each arm) and holograms. The first interferometer sorts between even and odd angular momenta. The sorted mode are then passed through a second stage where they are sorted further. In this second stage, the relative angle between the two Dove prisms is ${\pi/4}$. Even modes directly go through the interferometer, while odd modes go through an interferometer sandwiched between a unit-decreasing hologram and two unit-increasing holograms (one for each output arm). Once the modes are sorted according to their modes modulo $4$, desired operations on the first and the second qubits can be simultaneously performed. Moreover, we can apply an entangling operations, such as a controlled-not gate. Further studies would be required to include the measurement of the stabilizer.
Let us emphasize the peculiarity of the present approach with respect to two well known alternatives that might seem at first sight equivalent. To encode, say, two qubits in optical modes, one could use four orbital angular momenta or even four spatial modes. Mathematically, the three encodings are very different. Indeed, as already mentioned, the present encoding corresponds to a two-qubit subsystem, that is, ${\textrm{rotor}= \textrm{qubit} \otimes \textrm{qubit} \otimes \textrm{rotor}}$, while the encoding using four angular momenta is a two-qubit subspace, i.e., ${\textrm{rotor}=(\textrm{qubit} \otimes \textrm{qubit}) \oplus \textrm{rest}}$, and the four spatial modes form a qudit system, i.e., ${4=\textrm{qubit}\otimes \textrm{qubit}}$.

We now mention two other implementations. An alternative to photons is the use of one or many cold neutral atoms in a circular magneto-optical trap \cite{Sauer01,Morinaga08}. A magnetic field and several laser are used to create a ring-shaped potential where the atoms sit. Then the atoms are manipulated with laser pulses. Another option is concerned with ultra cold molecules, typically carbon-based molecules with a quantized rotational motion \cite{Medycki03,Mamone09}. Promising molecules are polar diatomic molecules trapped in an optical lattice or a solid matrix \cite{Krause04,Charron07}.

The three directions cited above are three possible options for implementing a rotor operating in the quantum regime. They would deserve additional investigations.
 
\section{Encoding qudits in a rotor} \label{discussion}
In Sec.~\ref{encoding} and \ref{protection} we have presented a systematic way to encode and protect $N$ qubits in a rotor. This construction can immediately be generalized to qudits. In this section we briefly summarize the relevant results.

We can define the $d$th root of unity ${\omega=e^{2 i \pi/d}}$, where $d$ stands for the dimension of the qudit. The Weyl pair ${(Z_j^{(d)},X_j^{(d)})}$, ${j=1,\cdots,N}$, where $N$ specifies the number of qudits, takes the form \cite{bergesbook}
\begin{align}
Z_j^{(d)}&=\omega^{\floor{L/(d^{j-1}(2\Delta L +1))}}, \nonumber \\
X_j^{(d)}&=V^{d^{j-1}(2\Delta L +1)}\nonumber \\
&-\big(\openone-V^{\dagger d^{j-1}(2\Delta L +1)}\big)P_{Z^{(d)}_j\!,1}V^{d^{j-1}(2\Delta L +1)},
\end{align}
where $P_{Z^{(d)}_j\!,1}$ is the projector onto the eigenspace of $Z_j^{(d)}$ with eigenvalue $+1$. One can check that we still have the required tensor product structure ${\textrm{rotor} = \textrm{qudits} \otimes (\textrm{error in $\ell$ }) \otimes \textrm{rotor}}$ between the encoded qudits, the errors in the angular momentum and the residual rotor. More specifically the Weyl pair of the residual rotor is ${\big(e^{i\alpha\lfloor L/(d^N(2\Delta L +1))\rfloor},V^{d^N(2\Delta L +1)}\big)}$ and the commutators between the Weyl pairs of the qubits, $S_L$ and the Weyl pair of the residual rotor vanish.

The stabilizer is ${S_{L}=e^{2i\pi n L/m}}$ and ${S_{\theta}=V^{m}}$, where now ${n=d^N}$ and ${m=d^N(2\Delta L +1)}$. The protected qudits are still given by Eq.~(\ref{codeword}) and Eq.~(\ref{solangularposition}).

\section{Conclusion}\label{conclusion}
We have presented a scheme to encode many genuine qubit subsystems in a rotor. We have shown how to manipulate and entangle them. We have also considered a quantum error-correcting code to protect the many qubits against small errors in angular position and momentum. The whole scheme is actually generalizable to qudits of any finite dimension. Furthermore we have considered physical approximations of the ideal scheme and their consecutive error probabilities. We have then turned to implementations and proposed several directions.

Theoretically, the advantage of encoding many qubits in a single degree of freedom is to allow their individual manipulation with a single unitary transformation. Thus complex entangled states like cluster states can in principle be realized with a single manipulation of the rotor. In this article, good care has been taken to encode genuine qubits subsystems in a rotor. A benefit is the freedom to use the remaining rotor to extract new qubits. Another advantage, inherent to the tensor product structure, is the possibility to correct errors. However the distinction between genuine qubits and pseudo qubits may become blurred after implementation. Nonetheless one might wonder whether genuine qubits and pseudo qubits exhibit a difference not only in terms of error model but also in operational terms. In other words, is there any task that can be performed with genuine qubits but that cannot be preformed equally well, or not at all, with pseudo qubits?

The challenges finally lie in the implementation. The scalability of the proposed scheme is directly related to the accuracy the rotor can be manipulated with. Therefore, with today's technology, it seems reasonable to reach a dozen of encoded qubits. This would already allow the simulation of small quantum systems. 

Let us conclude with a few comments on the orbital angular momentum of light. Unlike earlier proposals, the full continuous variable character of this degree of freedom is used, resulting in many genuine qubit subsystems. Moreover, qubits encoded in light are interesting not only because they are manipulated solely with linear optical elements but also because they are {\it flying} qubits. Thus one might conceive a computation taking place during the travel time between two remote places.
\vfill

\begin{acknowledgments}
Centre for Quantum Technologies is a Research Centre of Excellence funded by Ministry of Education and National Research Foundation of Singapore. J.S. is supported by MEXT and would like to thank B.-G.E. and Centre for Quantum Technologies for their kind hospitality. Ph.R. wishes to thank Kae Nemoto and the National Institute of Informatics for their warm hospitality. The authors wish to express their grateful feelings to Christian Kurtsiefer and Bill Munro for enlightening comments and friendly discussions. 
\end{acknowledgments}

\end{document}